\documentclass[12pt]{article}

\usepackage{newtxtext,newtxmath, url}

\usepackage{graphicx}

\usepackage[letterpaper,margin=1in]{geometry}

\renewenvironment{abstract}
	{\quotation}
	{\endquotation}

\date{}

\makeatletter
\renewcommand{\fnum@figure}{\textbf{Figure \thefigure}}
\renewcommand{\fnum@table}{\textbf{Table \thetable}}
\makeatother

\usepackage{scicite}

\usepackage{url}

\newcommand{\um}{\,\textmu m }	

\newcommand{\updated}[1]{#1}

\def\scititle{
	Neptune's Inner Moons and Rings Are Exposed Icy Body Interiors
}
\title{\bfseries \boldmath \scititle}

\author{
	M.~Ryleigh~Davis$^{1\ast, \updated{\dagger}}$,
    Matthew~Belyakov$^{1}$,
    Ian~Wong$^{2}$, \and
    Zachariah~Milby$^{1}$, 
    Michael~E.~Brown$^{1}$ \and
	\small$^{1}$Division of Geological and Planetary Sciences, California Institute of Technology, Pasadena, CA 91125, USA.\and
	\small$^{2}$Space Telescope Science Institute, Baltimore, MD 21218, USA\and
	\small$^\ast$Corresponding author. Email: \updated{rdavis1@ucsd.edu}\\
    \small\parbox{\textwidth}{\updated{$^\dagger$Current address: Department of 
    Astronomy \& Astrophysics, University of California San Diego, 
    La Jolla, CA 92093, USA.}}
}

\begin{document} 

\maketitle

\begin{abstract} \bfseries \boldmath
Neptune's single large moon, Triton, is accompanied by a set of dusty rings and small moons whose composition and origin are uncertain. Using the James Webb Space Telescope, we observed Neptune's rings and three moons: Larissa, Galatea, and Proteus. \updated{Remarkably, these moons and rings have spectra that are distinct from other outer solar system objects and show no evidence of water ice, despite hosting deep 3 \um OH absorption bands.} \updated{Additionally, Larissa, Galatea, and the rings show a} compositional signature unique among outer solar system bodies: a 2.72 \um absorption band diagnostic of magnesium-rich phyllosilicates – a signpost of extensive aqueous alteration. These minerals likely formed in the \updated{interior} of primordial satellites destroyed during Triton’s violent capture or a tidally shredded dwarf planet. Our findings suggest that Neptune’s present-day \updated{inner} moons \updated{and rings} reaccreted from this deep interior material and therefore uniquely access the interior composition of icy outer solar system differentiated bodies.
\end{abstract}



\subsection*{Introduction}
\noindent Neptune is the only giant planet in the solar system that lacks an ordered system of large regular satellites, suggesting that it underwent a unique evolutionary pathway. More than 99\% of the mass of Neptune's satellite system is dominated by Triton, with most of the remaining mass contained in a rich inner system of small close-in satellites and rings \cite{stone1989_Voyager2Encounter}. Explaining and understanding the system's unusual configuration can provide clues needed to reconstruct its history. Based on its retrograde and highly inclined orbit, Triton is likely a captured Kuiper belt object \updated{(KBO)} \cite{goldreich1989_NeptunesStory}. Had Neptune initially contained a system of large regular satellites, they would have been violently destroyed during Triton's capture, \updated{potentially} creating a debris disk that may have driven Triton's post-capture circularization \cite{goldreich1989_NeptunesStory, banfield1992_DynamicalHistoryInner, cuk2005_ConstraintsOrbitalEvolution, rufu2017_TritonsEvolutionPrimordial}. The fate of a primordial small inner satellite system following Triton's capture is uncertain; some models suggest that the small innermost moons may have survived Triton's capture or reaccreted from their own collisional debris, thus preserving much of their primordial composition \cite{banfield1992_DynamicalHistoryInner, karkoschka2003_SizesShapesAlbedos, li2020_OriginNeptunesUnusual}. Alternatively, Neptune's small inner moons may be the remnants of the originally larger regular satellites that were catastrophically disrupted \cite{cuk2005_ConstraintsOrbitalEvolution, rufu2017_TritonsEvolutionPrimordial}. Another possibility is that the inner moons and rings could be sourced from a passing KBO that was tidally disrupted within the planet's Roche limit sometime after Triton circularized \cite{dones1991_RecentCometaryOrigin, charnoz2009_DidSaturnsRings, crida2012_FormationRegularSatellites}. 

Understanding the formation history of Neptune's inner moon and ring system requires information on their compositions. Discovered only thanks to the 1989 Voyager 2 fly-by of Neptune \cite{stone1989_Voyager2Encounter}, the satellites are small and close to Neptune, making detailed study from the Earth difficult, and the compositions of these bodies and their associated rings remain largely unconstrained. \updated{Recent p}hotometric observations \updated{from the James Webb Space Telescope (JWST)} have revealed that the moons have a strong 3 \textmu m \updated{``O-H stretch'' absorption feature (hereafter referred to as the OH band)} that is weaker in the rings, interpreted as evidence for water ice \cite{belyakov2024_JWSTSpectrophotometrySmall, hedman2025_SpectralTrendsRings}. However, the satellites' low albedos ($<$10\%) and the survival of the innermost moons within Neptune's water ice Roche limit indicate that they must have a major contribution from dark rocky material with substantial internal strength \cite{karkoschka2003_SizesShapesAlbedos, tiscareno2013_COMPOSITIONSORIGINSOUTER, cuk2023sesquinary}. We obtained near-infrared spectra of the small inner moons---Proteus, Larissa, and Galatea---and faint rings to explore their composition and origin. We used the integral field unit (IFU) on JWST's Near-Infrared Spectrograph (NIRSpec) in PRISM mode (0.6--5.3 \textmu m at R$\sim$30--300). In the following sections, we discuss the unique composition of these moons as revealed by their spectra and the implications for the origin of the present-day Neptunian system. 

\updated{\subsection*{Results}}
The JWST/NIRSpec spectra of Neptune's rings and three of its inner moons: Proteus, Larissa, and Galatea, are shown in Fig. 1A. The surfaces of Neptune’s inner moons and rings are compositionally distinct from all other 
previously-observed outer solar system small bodies and moons. Proteus has a weak CO\textsubscript{2} absorption band at 4.27 \textmu m, which is marginally stronger on the trailing hemisphere than the leading, while Larissa, Galatea, and the rings show no evidence of CO\textsubscript{2}. \updated{All three moons and the rings} also lack a clear signature of water ice (e.g., 1.5, 1.65, 2.0, or 4.5 \um bands, or the 3.1 \um Fresnel peak), despite having very deep ($\sim$70\%) and broad 3 \um OH bands. \updated{The rings show a somewhat shallower 3 \um OH band depth and redder spectral slope relative to the three moons.}

\subsubsection*{Detection of Phyllosilicates}
Larissa, Galatea, and the ring spectra show an unusual additional absorption on the short-wavelength side of the broader 3 \textmu m OH band. The 3 \textmu m region of the Larissa, Galatea, \updated{and ring} spectra appears very similar to Proteus, except that Proteus either lacks or has a much weaker contribution from this additional absorption. Therefore, in order to isolate and characterize this feature, we \updated{remove the continuum by rescaling} all of the spectra to have the same band depth \updated{at 3.1 \textmu m}, then divide the spectra of Larissa, Galatea, and the rings by Proteus' spectrum. \updated{The continuum removed spectra shown in Fig. 1B} reveal a sharp checkmark-shaped 2.72-\textmu m absorption band that is characteristic of Mg-bearing \updated{serpentine-like} phyllosilicate \updated{clays} commonly seen on carbonaceous chondrites and main belt asteroids, which are associated with post-accretional aqueous alteration in their parent bodies \cite{zolensky1989_AqueousAlterationHydrous, takir2013_NatureDegreeAqueous}. No similar feature has ever been detected on any body in the outer solar system past Jupiter.   

We compare the 2.72-\textmu m band on the inner moons and rings with the best-fit meteorite spectrum available in the RELAB spectral library \cite{pieters2004relab} in Fig. 1B. The meteorite spectra were normalized and scaled to match the observed band depths. The 2.72-\textmu m feature is a remarkably good match to the phyllosilicate absorption band in the Meteorite Hills (MET) 00639, Allan Hills (ALH) 84031, and Cold Bokkeveld CM2 carbonaceous chondrite spectra.  

CM and CI carbonaceous chondrites are among the most primitive solar system materials that have been studied in the laboratory. They contain abundant phyllosilicates ($\gtrsim$60 vol\% for CM and $\gtrsim$80 vol\% for CI) which form during aqueous alteration of primary (unaltered) anhydrous silicates in their asteroid parent bodies \cite{tomeoka1985_IndicatorsAqueousAlteration, zolensky1997_CMChondritesExhibit, zolensky2008_RecordLowTemperatureAlteration,rubin2007_ProgressiveAqueousAlteration, howard2011_ModalMineralogyCM}. Laboratory studies show the exact band center and shape of the $\sim$2.7–2.8 \textmu m phyllosilicate feature trace the extent of aqueous alteration in the parent bodies of meteorites \cite{beck2010_HydrousMineralogyCM, takir2013_NatureDegreeAqueous}. Fig. 1C compares vacuum spectra, obtained at 150~K  \cite{takir2013_NatureDegreeAqueous}, of CM meteorites with varying degrees of alteration based on the Rubin petrologic scale \cite{rubin2007_ProgressiveAqueousAlteration}. The band evolves from a rounded 2.8-\textmu m feature (Fe-serpentine; e.g., QUE 97990) to a sharp 2.7-\textmu m ``checkmark" shape (Mg-serpentine; e.g., LAP 02277) with progressive alteration \cite{beck2010_HydrousMineralogyCM, takir2013_NatureDegreeAqueous, takir2019_3mmReflectanceSpectroscopya}. The inner Neptunian moons and rings are consistent with the Mg-serpentine dominated composition of the most aqueously-evolved CM chondrites (petrologic type CM2.0--2.1), indicating that these moons have undergone extensive and prolonged aqueous alteration. 

In order to contextualize the bands seen on Larissa and Galatea, Fig. 1C also shows globally-averaged spectra of the asteroid Bennu from \textit{OSIRIS-ReX} \cite{hamilton2019_EvidenceWidespreadHydrateda} and Ceres from \textit{Dawn} \cite{desanctis2015_AmmoniatedPhyllosilicatesLikelya} for comparison with two well-studied bodies with notable phyllosilicate absorptions. The band on the inner Neptunian satellites is a much stronger match in both band center and shape to Ceres. Ceres experienced strong and pervasive aqueous alteration, and its measured mineralogy and geochemistry are consistent with CM/CI-like carbonaceous chondrites \cite{mcsween2018_CarbonaceousChondritesAnalogs}. Observations of the 2.7-\textmu m phyllosilicate band on asteroids have been sparse due to atmospheric absorption, but recent JWST observations of the large ($\gtrsim$350 km) low-albedo outer main belt asteroids Hygiea and Pallas, along with Ceres, show that all three objects have similar Mg-rich and Fe-poor \updated{serpentine-like} phyllosilicates \cite{rivkin2025_ObservationsQuantitativeCompositional}. The advanced degree of aqueous alteration experienced by these bodies may be related to their size, with larger objects allowing convection of interior melt during (partial) internal differentiation which effectively increases the water-rock ratio \cite{mcsween2018_CarbonaceousChondritesAnalogs}. The phyllosilicate component of the inner Neptunian moons is broadly consistent with Ceres and the largest main belt asteroids, perhaps suggesting that they formed in differentiated parent bodies that are much larger than the moons' current sizes.


\updated{\subsection*{Discussion}}
\subsubsection*{Origin of the Phyllosilicates}
Larissa, Galatea, and the Neptunian rings contain abundant CM2-like Mg-endmember phyllosilicates that are formed by prolonged exposure ($\gtrsim$1--10Myr) to liquid water at moderately low temperatures ($\lesssim$300–400 K) \cite{zolensky1989_AqueousAlterationHydrous, deleuw2010_CarbonatesCMChondrites,fujiya2013_MnCrAges}. These aqueous conditions do not exist on any known outer solar system surfaces, and the formation of phyllosilicates requires both substantial and prolonged heating. Collisions are one potential mechanism, as laboratory experiments demonstrate that shock processes can generate localized phyllosilicates on water-rich surfaces \cite{furukawa2011_ImpactinducedPhyllosilicateFormationa} \updated{and post-impact hydrothermal alteration has been documented in craters with diameters as small as 20--30 km on terrestrial surfaces \cite{kirsimae2013_ImpactInducedHydrothermal, osinski2013impact, sapers2017evidence}}. However, impact-induced thermal metamorphism in chondrites is linked with dehydration of phyllosilicates and/or \updated{recrystallization} into olivine and pyroxene \cite{nakamura2006_Yamato793321CM, rubin2012_CollisionalFacilitationAqueous} \updated{and craters of 20--30 km diameter ($\sim$1/6 the diameter of Larissa and Galatea) achieve local temperature increases of only $\sim$100--200 K \cite{osinski2013impact}, insufficient to melt water ice at the $\sim$50 K surface temperatures of Neptune's inner moons. Additionally, post-impact hydrothermal systems within craters of this size persist on timescales orders of magnitude shorter than required to produce CM2-like phyllosilicates \cite{osinski2013impact}. An impactor capable of producing even this modest crater would likely be catastrophically destructive to bodies of this size, essentially ruling out a larger impactor capable of achieving the required $\Delta$T over sufficient timescales.} The production of the observed highly aqueously evolved Mg-serpentine end-member mineralogy via collisions is therefore unlikely.

In situ formation of phyllosilicates within moons as small as Larissa (\updated{$D\sim$}194 km) and Galatea (\updated{$D\sim$}176 km) is equally implausible due to thermodynamic constraints. Interior models for Uranus' moons suggest that under most formation conditions, Miranda (\updated{$D\sim$}470 km) will never generate enough internal heat to melt water ice and likely required a substantial contribution from tidal heating to achieve differentiation \cite{castillo-rogez2023_CompositionsInteriorStructures}. Thus, we can rule out endogenous alteration for Larissa and Galatea, which are approximately 2.5 times smaller than Miranda.

While aqueously altered phyllosilicates have never been detected in the outer solar system (beyond Jupiter, see \cite{sharkey2025jwst}), they are likely present in the interiors of many of the large \updated{moons} that have liquid water \cite{castillo-rogez2023_CompositionsInteriorStructures, tobie2014origin, sotin2021titan, zandanel2022geologically}, \updated{as well as} the interiors of large differentiated dwarf planets \cite{desch2009_ThermalEvolutionKuiper, emery2024_Tale3Dwarf}. The most plausible reason for the detection of phyllosilicates on Larissa and Galatea is that Neptune's current inner moons are reaccreted fragments containing \updated{aqueously altered} material from the deep interiors of \updated{one or more} differentiated icy bodies \updated{-- perhaps arising from a primordial system of regular satellites that previously orbited Neptune and were destroyed.}


\subsubsection*{Distinguishing between a regular satellite and Kuiper belt source for the inner satellites}

\updated{Two broad scenarios can deliver phyllosilicate-bearing interior material to Neptune's present inner moon system: the catastrophic destruction of Neptune's original regular satellite system during Triton's capture, or the tidal shredding of a large differentiated KBO within Neptune's Roche limit \cite{rufu2017_TritonsEvolutionPrimordial, crida2012_FormationRegularSatellites, hyodo2017_RingFormationGianta}}. \updated{However, companion JWST observations of Nereid — Neptune's third-largest moon, historically classified as an irregular satellite — provide independent evidence that a primordial regular satellite system existed around Neptune prior to Triton's capture \cite{belyakov2026_NeriedRegularSatellite}. Nereid's near-infrared spectrum is dominated by crystalline water ice and is inconsistent with the spectrum of any observed KBO or irregular satellite \cite{belyakov2026_NeriedRegularSatellite}. Combined with dynamical simulations demonstrating that Triton's inward migration could have perturbed an initially regular satellite onto Nereid's present eccentric orbit, this work suggests that Nereid is the sole surviving intact member of Neptune's original satellite system \cite{belyakov2026_NeriedRegularSatellite}, providing observational support for the existence and subsequent destruction of such a system during Triton's capture. We therefore favor the interpretation that the phyllosilicate-bearing material on Larissa, Galatea, and the rings originated from the interiors of these primordial regular satellites, rather than from an independent tidal disruption event involving a passing KBO, as this explanation requires no additional dynamical events beyond the minimum assumed components of the Neptunian system's history — namely, an original system of regular satellites, Triton's capture, and the subsequent collisional evolution of the debris disk — which are already invoked to explain the present-day configurations of Triton and Nereid.}

\updated{Nevertheless, both scenarios face the same difficulty in explaining the absence of water ice on the present-day inner moons.} Large regular satellites that melted water ice in their interiors are expected to have produced water-ice-rich outer shells, as is observed at Uranus \cite{castillo-rogez2023_CompositionsInteriorStructures}. Similarly, in all but the most extreme cases the material reaccreted from a tidally shredded KBO should contain a substantial fraction of water ice \cite{hyodo2017_RingFormationGianta}. It is therefore surprising that we find no evidence of water ice on any of the observed moons or rings and only small amounts of CO$_2$ on Proteus, suggesting that the reaccreted material was volatile poor or that these moons have lost most of their water ice during subsequent collisional processing. \updated{Indeed, the entire present-day inner satellite system of Neptune comprises only $\sim$1\% of the combined mass of Uranus's five large regular satellites (assuming a bulk density of 2 $g/cm
^{3}$), underscoring how much mass — and its associated water ice inventory — the disruption and collisional evolution of the debris disk must have removed.} Additionally, \updated{we note that} Proteus \updated{($D\sim$420 km)}, \updated{the largest and outermost of the inner satellites,} contains \updated{the same hydrated material that is responsible for the 3 \um OH band on the other moons and rings} but is phyllosilicate-poor, suggesting that it either reaccreted from hydrated material that never underwent aqueous alteration or subsequently reached temperatures high enough to dehydrate the phyllosilicate phases and weaken the characteristic 2.7-\textmu m absorption band \cite{che2011_SpectroscopicStudyDehydration}. \updated{It is possible that impacts, such as the one hypothesized to have ejected material from Proteus and formed the nearby small moon Hippocamp ($D\sim$35km) \cite{showalter2019_SeventhInnerMoon}, may have heated the surface of Proteus enough to dehydrate any phyllosilicates that were present. Alternatively, Proteus orbits beyond the synchronous rotation line, perhaps implying it reaccreted from material sourced from a different region of the satellite debris disk and/or from a different parent body than Larissa and Galatea.} \updated{The rings' shallower OH band depth and redder spectral slope relative to the moons may reflect differences in particle size rather than bulk composition — smaller particles preferentially suppress band depths while enhancing scattering at short wavelengths — though a contribution from compositional differences cannot be ruled out.} Detailed investigation into the \updated{destruction and} reaccretion properties of a satellite debris disk around Neptune, especially one whose lifetime may have been extended beyond typical reaccretion timescales due to resonant encounters with Triton \cite{zhang2008_OrbitalResonancesInner}, will be required to better understand the distribution of materials among the inner Neptunian moons.

An alternative explanation for the absence of water ice in the KBO shredding scenario is the disruption of a single object an order of magnitude more massive than Pluto, which simulations suggest could result in a nearly silicate-dominated reaccreted composition \cite{hyodo2017_RingFormationGianta}. In this scenario, anywhere from 0.1-10\% of the original KBO's mass may form a ring around Neptune, consistent with a single Pluto sized object sourcing Neptune's inner system \cite{hyodo2017_RingFormationGianta}. However, such an encounter and its initially massive debris disk would likely have destabilized Triton \cite{rufu2017_TritonsEvolutionPrimordial}. If the tidal shredding event occurred before Triton's capture, it is also possible that the system of original regular satellites discussed above could have formed out of material delivered by a disrupted dwarf planet, although this hybrid scenario is essentially indistinguishable except for the fact that the aqueous alteration may have occurred in the original dwarf planet rather than \updated{in the primordial} regular satellites themselves. A better understanding of the composition and temperature limits of \updated{the unidentified 3 \um} hydrated minerals \updated{present across the system} will likely be needed to more definitively distinguish between \updated{the tidally shredded KBO and destroyed primordial satellite scenarios, but the spectroscopic evidence from Nereid provides the most direct indication yet that the primordial satellite destruction pathway is the more natural explanation for the observed inner moon compositions.}

\subsubsection*{Source of the 3 \um Hydrated Material and the Question of Ammoniation}

\updated{All of the observed moons and rings share an unidentified hydrated mineral component that is responsible for the deep 3 \um OH band, for which no clear spectral analog exists in} laboratory spectral databases including the United States Geological Survey (USGS) \cite{kokaly2017_USGSSpectralLibrary}, Winnipeg Planetary Spectrophotometer Facility (PSF) \cite{cloutis2015mineral}, Keck/NASA Reflectance Experiment Laboratory (RELAB) \cite{pieters2004relab}, and Solid Spectroscopy Hosting Architecture of Databases and Expertise (SSHADE) \cite{schmitt2018solid}. Outer solar system objects with $\gtrsim$50\% 3 \textmu m band depths -- including water-rich KBOs \cite{pinilla-alonso2025_JWSTDiSCoTNOsPortrait, holler2025descriptive}, Neptune Trojans \cite{markwardt2025colors}, the regular satellites of Uranus \cite{brown1983uranian, cartwright2018_RedMaterialLarge}, and Saturnian irregular satellite Phoebe \cite{clark2012_SurfaceCompositionIapetus, belyakov2025_SaturnianIrregularSatellites} — all show additional spectral features associated with water and other volatile ices \updated{that are not present on Neptune's inner moons and rings}. Fig. 2A shows a comparison of Proteus' spectrum to other outer solar system objects known to have relatively deep 3 \textmu m bands without water ice signatures, with Proteus \updated{(and the other Neptunian moons and rings)} 70\% band \updated{depths} clearly being an outlier. \updated{Fine-grained water ice coatings have been proposed to explain shallow ($\sim$10\%) 3 \um bands on objects such as Themis \cite{rivkin2010_DetectionIceOrganics}, though this interpretation is no longer favored in the literature; regardless, such coatings cannot reproduce the $\sim$70\% band depths seen here and are therefore not viable.} \updated{Among objects shown in Fig. 2A,} an unidentified hydrated material is the favored explanation for the somewhat similar 3 \um absorption bands on Siarnaq and Albiorix as well as the unusual Neptune Trojan 2006 RJ103 \cite{belyakov2025_SaturnianIrregularSatellites, markwardt2025colors}. From direct spectroscopic comparison, it is uncertain whether the inner Neptunian satellites, 2006 RJ103, and small Saturnian irregular satellites share a common hydrated mineral, as the band shapes differ; \updated{the Neptunian bands are closest in shape to 2006 RJ103 \cite{markwardt2025colors}, though the Trojan lacks the 3.07 \um band minimum.}

\updated{The identity of the 3 \um hydrated component may carry important physical information about the original Neptunian satellites. One possibility is that it represents} relatively unaltered material from which the original satellites accreted — perhaps deriving from smaller, undifferentiated original satellites — and therefore probes the initial rocky component present in the outer solar system material that fed Neptune's protoplanetary disk. Alternatively, this material could come from the deepest and hottest regions of the interiors of the original satellites \updated{(or KBO)}, where \updated{the temperature/pressure conditions} were sufficient to metamorphically alter or partially dehydrate phyllosilicates \updated{or form a different hydrated phase}. In this case, identifying the composition of this material may offer key insights into the peak temperatures reached in the original satellites and therefore their sizes, as the maximum temperatures must have been lower than those required to fully dehydrate this phase. Acquisition of laboratory spectra of candidate materials under outer solar system conditions will be necessary to distinguish between these hypotheses.

All four spectra of Neptune's inner moons and rings show a potential weak feature near 3.07 \um that could plausibly be attributed to ammonia-bearing material (Fig. 2B), analogous to the ammoniated phyllosilicate band seen on Ceres \cite{desanctis2015_AmmoniatedPhyllosilicatesLikelya}. On Ceres and Hygiea, this feature has been interpreted as implying (1) formation in the outer Solar System where ammonia could condense followed by inward migration, or (2) accretion of ammonia-rich material (e.g. organics) that drifted inward from outer Solar System reservoirs that is only incorporated into clays in the largest objects that were heated sufficiently to release ammonia \cite{desanctis2015_AmmoniatedPhyllosilicatesLikelya,mcsween2018_CarbonaceousChondritesAnalogs}. However, the 3.07 \um feature on the Neptunian moons is most pronounced on Proteus\updated{ -- the moon that is most phyllosilicate-poor -- which argues against ammoniated phyllosilicates as the carrier.} A more likely interpretation is that ammonia remains contained within precursor outer solar system material and was incorporated into the unidentified hydrated phase responsible for the broader 3 \um band, consistent with an origin in a body large enough to melt water ice but which did not reach temperatures sufficient to release ammonia from its precursor organics.

\subsubsection*{Implications for the proto-Neptunian System}

The presence of phyllosilicates on Larissa and Galatea implies that at least one of Neptune's primordial moons (or precursor KBO) must have been large enough and experienced enough heating to melt ice in the interior. In contrast, these objects did not generate enough heat to completely dehydrate the \updated{ubiquitous} material that has maintained a high degree of hydration, as evidenced by the very deep 3 \textmu m OH band. Interior thermal models of Pluto suggest that it may have reached temperatures in excess of 1100 K \cite{robuchon2011_ThermalEvolutionPluto}, likely leading to the dehydration of \updated{the 3 \um} material and perhaps suggesting that aqueous alteration within smaller \updated{precursor} regular satellites is more likely. The presence of abundant phyllosilicates on Larissa and Galatea also requires that the bulk of this phyllosilicate-rich material has never been heated above $\sim$700K, including by impact shock, or the phyllosilicates would have dehydrated \cite{nakamura2006_Yamato793321CM, rubin2012_CollisionalFacilitationAqueous}. This means the relative speeds of the impacts that initially disrupted the primordial satellite system, as well as the subsequent collisional processing in the debris disk and after reaccretion, must have been low enough to avoid heating the bulk of the satellite material beyond these temperature limits. While the exact temperatures reached post-impact depend on numerous factors, such as the relative sizes and compositions of the impactors, the angle \updated{and speed} of impact, and the initial \updated{structure and} temperatures \updated{of the satellite interior(s)}, \updated{our} observationally derived temperature limits may inform models of the formation of Neptune's regular satellites, Triton's capture, and the subsequent evolution of the debris disk and Neptunian system.

\subsubsection*{Implications for Outer Solar System Icy Satellites}

The phyllosilicates present on the inner Neptunian moons require prolonged aqueous alteration, which must have occurred in the deep interior of a large differentiated object followed by their exposure in a catastrophic event. Our observations most readily support a scenario in which the original Neptunian satellite system was destroyed during Triton’s capture, and the current system reaccreted from the debris, thereby exposing the bulk core-mantle material of the original satellites. While aqueous alteration in the interiors of outer solar system icy bodies \updated{and moons} has long been predicted, this material typically remains buried deep in the interiors. The inner Neptunian moons are therefore the only place in the solar system where we can directly probe the interior of an icy satellite, or possibly a dwarf planet if the aqueous alteration occurred prior to the formation of Neptune's original system. In either case, the presence of CM2-like phyllosilicates on Neptune's inner moons demonstrates that, when heated sufficiently, outer solar system materials undergo a mineralogical evolution nearly identical to that observed in main-belt asteroids. This similarity implies that the rocky components present in the inner and outer solar system during planetesimal formation must have either been inherently similar (likely requiring thorough mixing of materials across the protosolar disk) or converged toward the same geochemical end state when aqueously altered despite initial differences in Fe/Mg ratios, olivine/pyroxene proportions, and chondrite abundances \cite{leone2023_IgneousProcessesSmall}.

\updated{\subsection*{Materials \& Methods}}

\subsubsection*{Observations and Data Reduction}

Proteus, Larissa, and Galatea were observed with JWST NIRSpec as part of the Cycle 3 GO Program \#4645. The observations were obtained using the Integral field unit (IFU) in PRISM mode (0.6-5.3~{\textmu}m,  R$\sim$30--300). Each satellite was observed multiple times and the resulting spectra were combined to achieve the highest possible S/N. The details for each observation are listed in Table S1. The total coadded integration times for Proteus, Larissa, and Galatea were 2334.4s, 5514.6s, and 8169.6s, respectively. The ring spectra were combined from all of the observations of Larissa and all but the last dither in each set of Galatea observations, with a total integration time of 11,641.8s.

Due to the substantial scattered light from Neptune present in all of our observations, we implemented a custom reduction of the JWST data that combines the standard pipeline with our own background subtraction\updated{, as detailed below,} and extraction of the data. We used version 1.18.0 of the JWST pipeline with \texttt{jwst\_1364.pmap} for calibration. We begin with the raw ``.uncal'' files, and run the standard Stage 1 and 2 pipeline, generating IFU cubes aligned with the detector. For Galatea and Larissa, we run the pipeline twice to generate both detector and sky-aligned data cubes.

To extract the moons Proteus, Galatea, and Larissa, we begin by masking outliers in the data cube, removing values below 0, and masking any values greater than maximum flux at the center pixel of each moon across all wavelength slices. We find this more robust than aggressive sigma-clipping, which can mask pixels with substantial scattered light from Neptune. For background subtraction, we mask the target in a radius of 3 pixels, then interpolate over the masked pixels using a two-dimensional Gaussian kernel with 1.5 pixel width to obtain a rough estimate of the background. We then subtract the background from the image, and perform PSF extraction as described in \cite{brown2023_StateCOCO2, wong2024_JWSTNearinfraredSpectroscopy}. We similarly extract a solar analog spectrum from program 1128 using a 3 pixel aperture, and divide our extracted spectra by the solar analog to obtain a reflectance spectrum.

To extract the spectra of the rings, we began by determining their sky coordinates in J2000.0 right ascension and declination at the time of each of the NIRSpec observations using the SPICE toolkit \cite{Acton1996}. We assumed each ring was circular and had zero inclination. We used the semimajor axes for each ring determined during the Voyager 2 flyby on August 18, 1989 at 12:00 ET \cite{Porco1995}. We calculated the ring positions at $0.1^\circ$ intervals in the body-fixed \url{IAU_Neptune} frame then transformed them to the J2000 inertial frame. We calculated the right ascension and declination of each ring position using the SPICE routine \texttt{recrad}, then transformed those sky coordinates to pixel coordinates using the \url{skycoord_to_pixel} function available in Astropy's \texttt{wcs} module \cite{AstropyCollaboration2013,AstropyCollaboration2018,AstropyCollaboration2022}. 

Using now the sky-aligned cubes described above for Larissa and Galatea, we obtain pixel coordinates from the ra/dec coordinates of the Adams, Arago, and Le Verrier ring. The locations of the rings derived for two IFU images of Larissa are shown in Fig. S1, demonstrating the accuracy of the extracted coordinates. We mask out pixels that fall within 2.5 pixels of any of the rings or Larissa/Galatea, and interpolate with a two-dimensional gaussian kernel across the rings to estimate the background. We then subtract off the background, and find the median of each wavelength slice. The choice of specific parameters used in the reduction can cause major changes in the data below 1.7 microns and past 4.5 microns, where scattered light from Neptune dominates the spectrum. We therefore do not show the data in those regions. The region between 1.7-4.5 shows a similar spectrum regardless of choice of kernel width used for smoothing and area used to measure the spectra of the rings. We select a kernel width of 1.5 pixels because it results in the visually cleanest extraction.

Raw data from all of the observations are available at the Mikulski Archive for Space Telescopes (MAST) portal under program ID 4645. The code used to generate the reduced Stage 2 background subtracted files and extracted spectra is archived at \url{https://data.caltech.edu/records/vwdse-4d488} \cite{reduction_code}.

\begin{figure} 
	\centering
	\includegraphics[width=0.8\textwidth]{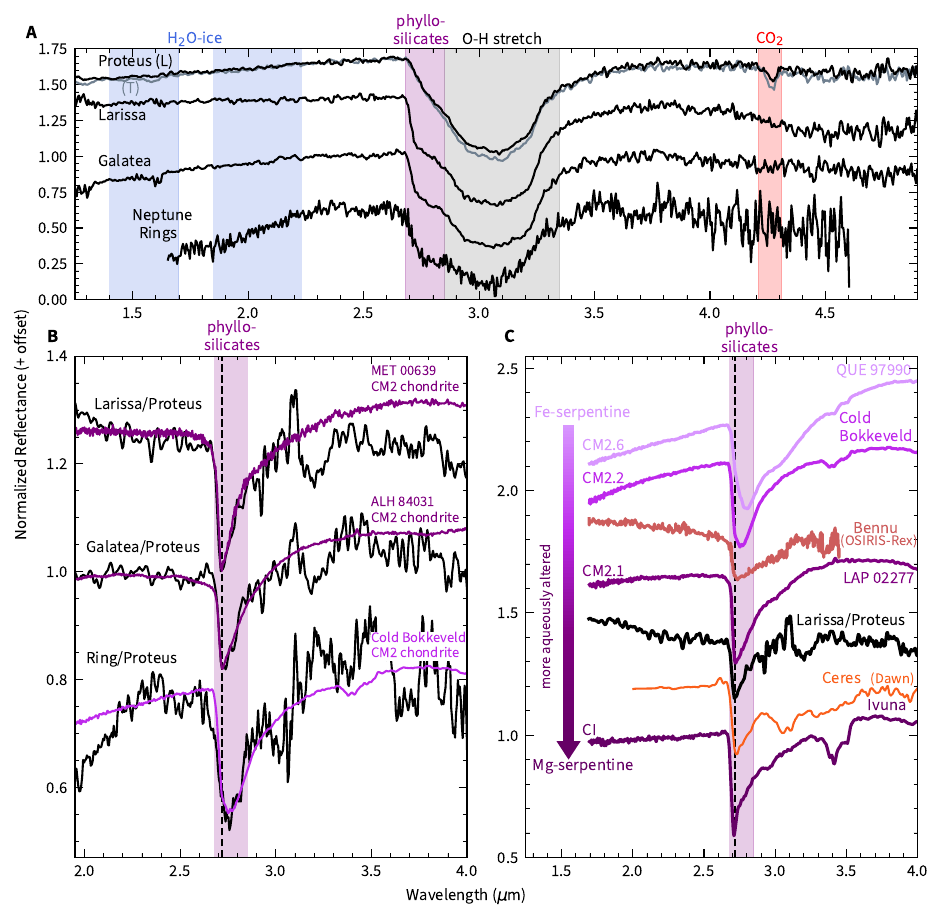} 

	\caption{\textbf{JWST/NIRSpec spectra of the inner Neptunian satellites reveal aqueously altered phyllosilicates from the deep interior of an original regular satellite (or KBO).}
	(\textbf{A}) Normalized reflectance spectra of Proteus, Larissa, Galatea and the average ring spectrum. Key absorption bands are highlighted, demonstrating the deep 3 \textmu m OH band and 2.7 \textmu m feature, CO$_2$ only on Proteus, and no water-ice overtones. (\textbf{B}) The continuum removed 2.72 \textmu m absorption band compared with meteorite laboratory spectra from the RELAB spectral library (purple). The band is a close match to CM2 carbonaceous chondrites. (\textbf{C}) Comparison with average spectra of asteroids' Bennu \cite{hamilton2019_EvidenceWidespreadHydrateda} and Ceres \cite{desanctis2015_AmmoniatedPhyllosilicatesLikelya} along with laboratory spectra of CM2 chondrites acquired under vacuum at 150 K (purple \cite{takir2013_NatureDegreeAqueous}) in order of increasing aqueous alteration as indicated by the Rubin petrologic scale \cite{rubin2007_ProgressiveAqueousAlteration}. The spectra range from the least altered Fe-serpentine dominated QUE 97990 to the most altered Mg-serpentine dominated LAP 02277 and CI chondrite Ivuna. The band on the inner Neptunian moons best matches the most altered CM2 chondrites and the largest asteroids, like Ceres.}
	\label{fig:spectra} 
\end{figure}

\begin{figure} 
	\centering
	\includegraphics{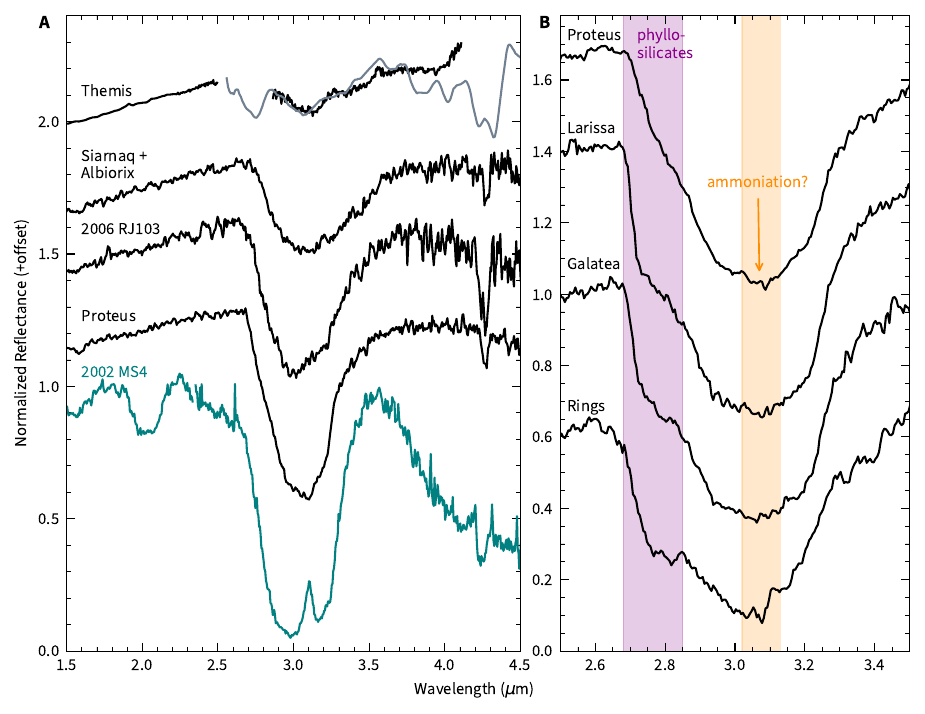} 
	\caption{\textbf{Comparison of Proteus to Outer Solar System bodies with 3 \textmu m OH bands \updated{and Possible Ammoniation Feature.}}
		\updated{(\textbf{A}) Neptune's inner satellites and rings} shows an unusually deep 3 \textmu m OH band among objects that do not show clear spectroscopic evidence of water-ice, like the Neptune Trojan 2006 RJ103 \cite{markwardt2025colors}, the average spectrum of the Saturnian irregular satellites Siarnaq and Albiorix \cite{belyakov2025_SaturnianIrregularSatellites}, and the main-belt asteroid Themis \cite{rivkin2010_DetectionIceOrganics, usui2019_AKARIIRCNearinfrared} shown here. Indeed, the band on the inner moons is almost as deep as 2002 MS4, a water-ice rich KBO with one of the deepest water bands in the observed population \cite{pinilla-alonso2025_JWSTDiSCoTNOsPortrait}. \updated{(\textbf{B}) In addition to the 2.72 \um phyllosilicate absorption, Neptune's inner moons and rings show a possible band near 3.07 \textmu m, broadly consistent with those attributed to ammoniation on Ceres, that is most apparent on Proteus.}
        }
	\label{fig:proteus3um} 
\end{figure}



\clearpage 

%
\bibliography{references} 
\bibliographystyle{sciencemag}

%
%
%
%
%
%


\section*{Acknowledgments}
The authors thank Dr. Driss Takir for providing laboratory data used in Fig. 1. M.R.D. thanks Dr. Josh Emery, Dr. Bethany Ehlmann, and Dr. Samantha Trumbo for helpful discussions during the preparation of this manuscript. This study was based on observations made with the NASA/ESA/CSA James Webb Space Telescope (JWST). The data were obtained from the Mikulski Archive for Space Telescopes (MAST) at the Space Telescope Science Institute, which is operated by the Association of Universities for Research in Astronomy under NASA contract NAS 5-03127 for JWST. 

\paragraph*{Funding:}
The observations analyzed here are associated with JWST program \#4645. Support for program \#4645 was provided by NASA through a grant from the Space Telescope Science Institute, which is operated by the Association of Universities for Research in Astronomy, Inc., under NASA contract NAS 5-03127.

\paragraph*{Author contributions:}
\hspace{5pt}\\
\updated{
Conceptualization: MRD, MB, IW\\
Data curation: MB\\
Formal analysis: MRD, MB\\
Investigation: MRD, MB\\
Methodology: MRD, MB, IW, ZM, MEB\\
Project administration: MRD, MEB\\
Resources: MB\\
Software: MRD, MB, IW, ZM\\
Supervision: MEB\\
Validation: MRD, MB\\
Visualization: MRD\\
Writing—original draft: MRD\\
Writing—review \& editing: MRD, MB, IW, ZM, MEB}

\paragraph*{Competing interests:}
The authors declare they have no competing interest.

\updated{\paragraph*{Data, Code, and Materials Availability:}
All data and code needed to evaluate and reproduce the results in the paper are present in the paper and/or the Supplementary Materials. The JWST data are available at MAST (\url{https://mast.stsci.edu/portal/Mashup/Clients/Mast/Portal.html}) under program ID 4645. The specific observations shown in this paper can be found at doi:10.17909/1cqh-hy49 \cite{jwstdoi}. The code used to generate the reduced Stage 2 background subtracted files and the extracted spectra is archived at CaltechData (\url{https://data.caltech.edu/records/vwdse-4d488}) \cite{reduction_code}. We have also used data from JWST program ID 1128 for calibration with a solar-like spectrum; however, use of any available solar analog spectrum is sufficient. The laboratory spectra used for comparison in Figure 1 are published in \cite{takir2013_NatureDegreeAqueous} or available online in the RELAB spectral library \cite{pieters2004relab} under sample ID's MP-TXH-226-D and MP-TXH-194-P. The spectra of Bennu and Ceres shown in Fig. 1C are published in \cite{hamilton2019_EvidenceWidespreadHydrateda} and \cite{desanctis2015_AmmoniatedPhyllosilicatesLikelya}. The JWST spectrum of 2006 RJ103 shown in Figure 2 for comparison is available at MAST under program ID 2550 and published in \cite{markwardt2025colors}, and the spectrum of 2002 MS4 is available under program ID 1191. Individual spectra of the Saturnian irregular satellites Siarnaq and Albiorix are available on MAST under program ID 3716 and the average spectrum shown in Figure 2 is published in \cite{belyakov2025_SaturnianIrregularSatellites}. The two Themis spectra shown are from \cite{rivkin2010_DetectionIceOrganics} and \cite{usui2019_AKARIIRCNearinfrared}. This study did not generate new materials.}


\subsection*{Supplementary materials}
Fig. S1\\
Table S1\\


\newpage


\renewcommand{\thefigure}{S\arabic{figure}}
\renewcommand{\thetable}{S\arabic{table}}
\renewcommand{\theequation}{S\arabic{equation}}
\renewcommand{\thepage}{S\arabic{page}}
\setcounter{figure}{0}
\setcounter{table}{0}
\setcounter{equation}{0}
\setcounter{page}{1} 


\begin{center}
\section*{Supplementary Materials for\\ \scititle}

M. Ryleigh Davis$^\ast$,
Matthew Belyakov,
Ian Wong,
Zachariah Milby
and Michael E. Brown\\ 
\small$^\ast$Corresponding author. Email: rdavis1@ucsd.edu\\
\end{center}

\subsubsection*{This PDF file includes:}
Fig. S1\\
Table S1\\


\newpage





\begin{figure} 
	\centering
	\includegraphics[width=\textwidth]{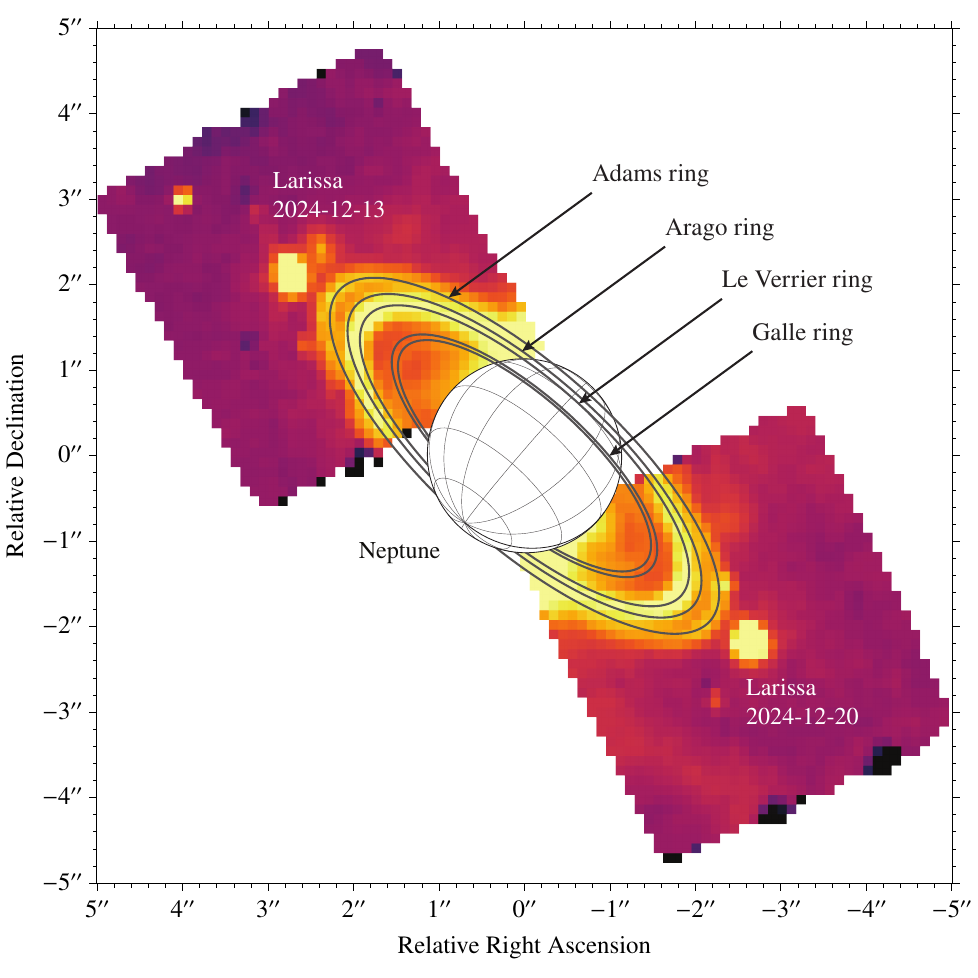} 
	\caption{\textbf{Representative JWST/NIRSpec IFU images of Larissa on 2024-12-13 and 2024-12-20 showing the locations of Neptune's rings.} The IFU images were created from wavelength slices between 2.1-2.35 \textmu m and acquired when Larissa was on opposite sides of Neptune. The bright Adams and Le Verrier rings, which are shepherded by Galatea and Despina, respectively, are easily visible in both images. The calculated ring coordinates are overlain in black.
        }
	\label{fig:sup_ifu} 
\end{figure}

\begin{table} 
	\centering
	\caption{\textbf{Details of JWST Observations of the Inner Satellites of Neptune from program \#4645.}
		Each individual observation is listed with the target, date in UTC, total exposure time in seconds, start and end time in UTC, and central longitude ($^\circ$W) and latitude ($^\circ$N/S) of the satellite in the middle of the observation. All observations were acquired using NIRSpec/IFU in PRISM mode with the clear filter.}
	\label{tab:observations} 

	\begin{tabular}{lccccr} 
		\\
		\hline
		Target & Date & Exposure Time & Start/End Time (UTC) & Obs Sub-Lon & Obs Sub-Lat\\
		\hline
		Larissa & 2024-12-13 & 919.1 s & 16:15:24/16:30:58 & 55$^\circ$W & 21$^\circ$S\\
		Larissa & 2024-12-13 & 919.1 s & 16:33:53/16:49:26 & 63$^\circ$W & 21$^\circ$S\\
        Larissa & 2024-12-13 & 919.1 s & 16:52:07/17:07:41 & 72$^\circ$W & 21$^\circ$S\\
        Proteus & 2024-12-13 & 291.8 s & 17:40:01/17:45:07 & 254.5$^\circ$W & 24.5$^\circ$S\\
        Proteus & 2024-12-13 & 291.8 s & 17:47:48/17:52:54 & 256$^\circ$W & 24.5$^\circ$S\\
        Proteus & 2024-12-13 & 291.8 s & 17:55:49/18:00:56 & 257.5$^\circ$W & 24.5$^\circ$S\\
        Proteus & 2024-12-13 & 291.8 s & 18:03:36/18:08:43 & 259.5$^\circ$W & 24.5$^\circ$S\\
        Proteus & 2024-12-16 & 291.8 s & 13:02:12/13:07:19 & 74.5$^\circ$W & 24.5$^\circ$S\\
        Proteus & 2024-12-16 & 291.8 s & 13:09:59/13:15:06 & 76.5$^\circ$W & 24.5$^\circ$S\\
        Proteus & 2024-12-16 & 291.8 s & 13:18:01/13:23:07 & 78$^\circ$W & 24.5$^\circ$S\\
        Proteus & 2024-12-16 & 291.8 s & 13:25:48/13:30:54 & 80$^\circ$W & 24.5$^\circ$S\\
        Galatea & 2024-12-17 & 1021.2 s & 15:33:55/15:51:11 & 226$^\circ$W & 20.9$^\circ$S\\
        Galatea & 2024-12-17 & 1021.2 s & 15:53:51/16:11:07 & 238$^\circ$W & 20.9$^\circ$S\\
        Galatea & 2024-12-17 & 1021.2 s & 16:13:48/16:31:03 & 245$^\circ$W & 20.9$^\circ$S\\
        Galatea & 2024-12-17 & 1021.2 s & 16:33:44/16:51:00 & 261$^\circ$W & 20.9$^\circ$S\\
        Galatea & 2024-12-19 & 1021.2 s & 08:44:32/09:01:48 & 226.5$^\circ$W & 20.9$^\circ$S\\
        Galatea & 2024-12-19 & 1021.2 s & 09:04:28/09:21:44 & 238$^\circ$W & 20.9$^\circ$S\\
        Galatea & 2024-12-19 & 1021.2 s & 09:24:24/09:41:40 & 250$^\circ$W & 20.9$^\circ$S\\
        Galatea & 2024-12-19 & 1021.2 s & 09:44:35/10:01:51 & 261.5$^\circ$W & 20.9$^\circ$S\\
        Larissa & 2024-12-20 & 919.1 s & 01:22:28/01:38:02 & 236$^\circ$W & 21$^\circ$S\\
        Larissa & 2024-12-20 & 919.1 s & 01:40:43/01:56:16 & 244$^\circ$W & 21$^\circ$S\\
        Larissa & 2024-12-20 & 919.1 s & 01:58:57/02:14:31 & 252$^\circ$W & 21$^\circ$S\\
		\hline
	\end{tabular}
\end{table}


\clearpage 





\end{document}